\documentclass[12pt]{article}

\columnwidth\textwidth

\begin{document}

\newcommand{\be}{\begin{equation}}
\newcommand{\ee}{\end{equation}}
\newcommand{\beann}{\begin{eqnarray*}}
\newcommand{\eeann}{\end{eqnarray*}}
\newcommand{\bea}{\begin{eqnarray}}
\newcommand{\eea}{\end{eqnarray}}
\newcommand{\nn}{\nonumber}
\newcommand{\ben}{\begin{enumerate}}
\newcommand{\een}{\end{enumerate}}
\newtheorem{df}{Definition}
\newtheorem{thm}{Theorem}
\newtheorem{lem}{Lemma}
\newtheorem{prop}{Proposition}
\begin{titlepage}

\noindent
\hspace*{11cm} BUTP-96/26 \\
\vspace*{1cm}
\begin{center}
{\LARGE Time evolution of observable properties of
reparametrization-invariant systems}

\vspace{2cm}

P. H\'{a}j\'{\i}\v{c}ek \\
Institute for Theoretical Physics \\
University of Bern \\
Sidlerstrasse 5, CH-3012 Bern, Switzerland \\
\vspace*{2cm}

December 1996 \\ \vspace*{1cm}

\nopagebreak[4]

\begin{abstract}
A short review of some recent work on the problem of time and of observables 
for the
re\-pa\-ra\-me\-tri\-za\-tion-invariant systems is given. A talk presented at
the Second Meeting on Constrained Dynamics and Quantum Gravity at Santa
Marguerita Ligure, September 17--21 1996.
\end{abstract}

\end{center}

\end{titlepage}

\section{Introduction}
\label{sec:intro}
In the quest for quantum gravity, a number of problems have been
recognized. This paper concerns mainly the problem of time and the
problem of observables \cite{isham} \cite{KK5} within the canonical
approach. 

The problem of time has many aspects; we shall focus on the time
evolution. In the quantum theory, the evolution is defined by the
Schr\"{o}dinger equation (or Heisenberg equations) and the basic
structure is the Hamiltonian operator. The classical counterpart
thereof is a Hamiltonian dynamical system \cite{AM} and there will be
a Hamiltonian function. In the classical theory of gravity---the
general relativity---we do not know what could play the role of
Hamiltonian: the general relativity is not a Hamiltonian dynamical,
but rather a reparametrization invariant system (RIS). In order to
capture the structure which underlies the Hamiltonian in general, we
give a definition of RIS such that all Hamiltonian dynamical systems
are included. Any Hamiltonian dynamical system can be rewriten in the
form of RIS by the well-known process of parametrization. It turns out
that the Hamiltonian dynamics is equivalent to a kind of reference
system in the phase space, the so-called auxiliary rest frame
(ARF). Each parametrization defines a unique ARF in the resulting RIS
and each choice of ARF in a RIS enables one to reformulate the RIS
uniquely as a Hamiltonian dynamical system \cite{PH1}.

The notion of ARF allows broad generalization such that RIS's
with a generalized ARF can still be brought to the form of a
(generalized) Hamiltonian dynamical system \cite{PH5}. Many apparently
different approaches to the quantum gravity can then be recognized as
just different choices of ARF \cite{PH4}. The time evolution between
two non global transversal surfaces can also be incorporated, see
\cite{HI2} (Hawking effect), \cite{PH2}, and \cite{PH5}.

However, the general relativity does not seem to possess any natural
(generalized) ARF. Thus, the question could arise: are there any
alternatives to the Schr\"{o}dinger equation form of dynamics in the
quantum theory? The answer is yes; an example of such a quantum theory
is the quantum field theory (QFT) on fixed curved background spacetime, if
this background does not admit any timelike Killing vector. We analyze
this model and describe the two strategies which were invented long
ago to cope with the problem: the scattering and the algebraic
methods \cite{HI2}. The dynamical equation that replaces the Schr\"{o}dinger
equantion in this case is the quantum field equation.

The problem of observables is closely related to that of time
evolution. An observable in a Hamiltonian dynamical system represents
a whole set of measurements, each perfomable at a particular time;
these measurements are considered to be ``the same'' (for example, the same
apparatus measures the position at different times) \cite{PH5}. It
turns out that a precise definition of such a ``standard'' observable
requires the same reference frame as the definition of time
evolution. Thus, only if an ARF is chosen, standard observables are
well-defined.

In general, the observable properties of RIS's can be represented by
re\-pa\-ra\-me\-triz\-a\-tion invariant functions, the so-called perennials.
Originally, perennials were called ``first-class quantities'' or
``observables'' by Bergmann and Dirac \cite{bergmann1},
\cite{dirac1}. Bergmann recognized the two major problems:
\ben
\item Are there enough perennials available? On one hand, even if they 
exist, they are difficult to find in practice. On the other, the very
existence of perennials has been questioned recently (cf.\ \cite{KK2} or
\cite{AT}).
\item Reparametrizations can be identified with diffeomorphisms that act
transitively along dynamical trajectories of RIS's. This implies that
perennials must be functions that are constant along classical
solutions (this is a justification for the name, which has been
introduced by Kucha\v{r} \cite{KK2} to distinguish them from standard
observables). In a theory, in which the perennials are used to
describe observable properties, no time evolution seems possible (the
``frozen dynamics'' of Bergmann).
\een

A new impetus for research in this area came from Ashtekar's \cite{bluebook}
work on canonical quantization. The papers \cite{R-group}, \cite{R-observ},
\cite{rovelli} by Rovelli and \cite{KK2} by Kucha\v{r} contain crucial
ideas. These ideas have been worked out in a rigorous and systematic way in
\cite{PH0}, \cite{PH1}, \cite{PH2}, \cite{HI1}, \cite{HI2} and \cite{PH5}.
Bergmann's problems have been solved or circumvented; in particular,
the existence of a sufficient number of perennials has been shown
\cite{PH5} for physically reasonable RIS's.

It turns out, then that a standard observable can be identified with a
certain class of perennials. The class can be defined by one perennial
and the ARF. The time evolution of the corresponding measurable
property is determined by the Schr\"{o}dinger or Heisenberg
equations. In the case that there is no ARF, one can still describe
the observable properties by a special kind of perennials. They have
been introduced by Rovelli, and we call them ``labelled perennials.''

The plan of the paper is as follows. In Sec.\ 2, we give a definition
of RIS that is sufficiently general for our purposes. The rest of the
section is devoted to the first problem of Bergmann. We restrict
ourselves to the RIS's that possess a manifold-like space of physical
degrees of freedom and show that they admit a complete system of
perennials. Then, we explain the general structure and physical
meaning of perennials. In Sec.\ 3, we describe the relation between
the Hamiltonian dynamics and the ARF. The physical meaning of the ARF
is analyzed; the definition of the standard observable follows
immediately from this analysis. The (classical) dynamical equations of
Schr\"{o}dinger and Heisenberg are then derived from the ARF. The
second problem of Bergmann disappears. In Sec.\ 4, we
describe examples of generalized ARF's. We show how some apparently
qualitatively different approaches to quantum gravity result just by different
choices of ARF. In Sec.\ 5, we study the parametrized QFT on fixed
classical background spacetime as an example of 
RIS that, on one hand, is relatively well-understood and that, on the
other, is afflicted with problems associated with a choice of ARF. We
review the well-known scattering and algebraic approaches to the
dynamics of this model. After everything is translated into the
language of RIS and perennials
\cite{HI1}, \cite{HI2}, these approaches can be seen as ARF-free methods
of describing dynamics; they are, in fact, using a particular kind of labelled
perennials.

\section{Existence and nature of perennials}
\label{sec:perenn}
The existence is mainly a mathematical problem, so we have to be relatively
precise.

Let $(\tilde{\Gamma},\tilde{\Omega})$ be a symplectic manifold; it will play
the role of the extended phase space. Let $\Gamma$ be a submanifold of
$\tilde{\Gamma}$. Let us define, for any $x \in \Gamma$, $\tilde{L}_x :=
\{X \in T_x\tilde{\Gamma}\,|\,\tilde{\Omega}(X,Y) = 0\ \forall Y \in
T_x\Gamma\}$ and the space $L_x$ of longitudinal vectors at $x$ by $L_x :=
\tilde{L}_x \cap T_x\Gamma$. Suppose that $L_x \neq \{0\}$ and that $L_x$ for
all $x \in \Gamma$ defines a subbundle $L\Gamma$ of the tangent bundle
$T\Gamma$. Then, the triplet $(\tilde{\Gamma},\tilde{\Omega},\Gamma)$ is called
\textit{reparametrization-invariant (constraint) system} (RIS) and $\Gamma$ is
called the \textit{constraint surface}; the points of $\Gamma$ are interpreted
as physical states.  $L_x$ is also the subspace of degeneracy of the pull-back
$\Omega$ of $\tilde{\Omega}$ to $\Gamma$. $L\Gamma$ is an integrable
subbundle, because $\Omega$ is closed. Thus, there are integral submanifolds
of $L\Gamma$ through any point $x$ of $\Gamma$. Let us call \textit{maximal}
integral submanifolds of $L\Gamma$ \textit{c-orbits} (orbits of first-class
constraints). For any RIS, there is one-to-one correspondence between 
c-orbits and maximal dynamical trajectories \cite{PH1}.  That means that each
classical solution is just a c-orbit rather that a set of c-orbits like, e.g.,
in gauge theories. The motivation for choosing the name
``reparametrization-invariant'' is that the dynamical trajectories of the
system are submanifolds (sets) rather than maps (like curves) and the
coordinates along such submanifolds (parametrization) is arbitrary.

Non-geometrical formulations of RIS's usually focuse on the
Hamiltonian action $S = \sum p\dot{q} - \mathcal{H}$, where the second
class constraints have been removed and $\mathcal{H}$ is a linear
combination of the first class ones (see \cite{dirac1}). Let us call
$\mathcal{H}$ \textit{extended Hamiltonian} to distinguish it from the
genuine (Schr\"{o}dinger or Heisenberg) Hamiltonian (see Sec.\
\ref{sec:H-S}), which will be called simply Hamiltonian.

The reparametrization-invariant systems form a very general class: they
include all (constrained) Hamiltonian dynamical systems, because such systems
can be transformed to equivalent RIS's by parametrization (see the example at
the end of this section). This holds also for the asymptotically flat sector
of general relativity, which can be parametrized at infinity (for an example
see \cite{KK1}). A particular class of RIS's is formed by the so-called
\textit{first-class systems}; they are defined by $L_x = \tilde{L}_x$ for all
$x \in \Gamma$; that means that $\Omega$ is maximally degenerate.

Consider the space $\bar{\Gamma}$ defined by $\bar{\Gamma} :=
\Gamma/\gamma$, where $\gamma$ denotes the c-orbits. With the quotient
topology, $\bar{\Gamma}$ is a topological space; it is interpreted as the
space of physical degrees of freedom (the number of physical degrees of
freedom is half of the dimension of $\bar{\Gamma}$), or the space of classical
solutions, or the physical phase space. An important notion in the theory of
RIS's is that of \textit{transversal surface} (a section of
$\Gamma/\gamma$). It is any submanifold of $\Gamma$ which has no common
tangent vectors with any c-orbit (except for zero vectors) and which
intersects each c-orbit in at most one point. A \textit{global transversal
surface} must intersect every c-orbit.

Our formal mathematical definition implies little restriction on the topology
of $\bar{\Gamma}$. Consider the following example (an ergodic system, cf.\
\cite{AA}). Let $(M,g)$ be a compact Riemannian
(positive-definite metric) manifold with negative constant curvature. Let
$\tilde{\Gamma}$ be $T^*M$ with coordinates $x^\mu, p_\mu$ and $\tilde{\Omega}$
be the standard symplectic form of cotangent bundles. Let $\Gamma$ be given by
the equation $g^{\mu\nu}p_\mu p_\nu = 1$. The dynamical trajectories of this
RIS are geogesic arcs of $(M,g)$. One can show that $\bar{\Gamma}$ has the
coarsest possible topology: just the empty set and $\bar{\Gamma}$ itself are
open; there is no transversal surface through any point of $\Gamma$.

However, for a RIS to be physically sensible, its physical degrees of freedom
must form, at least locally, a manifold. For general relativity, the structure
of this space has been studied for some time. Large portions of it have been
proved to possess manifold structure; points, at which the manifold structure
breaks down have also been found, but they form a small subset (at least as
yet, a recent paper is \cite{FM}). That motivates the following definition
\cite{PH5}:
\begin{df}
A RIS $(\tilde{\Gamma},\tilde{\Omega}, \Gamma)$ is called \textit{locally
reducible}, if an open dense subset of $\bar{\Gamma}$ is a quotient manifold.
\end{df}

For the sake of simplicity, we assume that $\bar{\Gamma}$ is a quotient
manifold; the general case is dealth with in \cite{PH5}. Then, the natural
projection $\bar{\pi} : \Gamma \mapsto \bar{\Gamma}$ is a submersion and it
determines a two-form $\bar{\Omega}$ by $\bar{\pi}^*\bar{\Omega} =
\Omega$. The symplectic manifold $(\bar{\Gamma},\bar{\Omega})$ is the physical
phase space. $\bar{\pi}^*$ also gives a one-to-one relation between functions
on $\bar{\Gamma}$ and functions on $\Gamma$ that are constant along all
c-orbits. We define: a function $o : \tilde{\Gamma} \mapsto
\mathbf{R}$ that is constant along all c-orbits is a
\textit{perennial}. We call a set of perennials \textit{complete}, if
it separates points in an open dense subset of $\bar{\Gamma}$ (cf.\
\cite{bergmann1}).

Now, the example above does not admit any perennial. There are some results on
an analogous problem from the theory of Hamiltonian systems suggesting that
the existence of perennials for a generic RIS is very unlikely: a theorem by
Robinson \cite{robinson}. It states, roughly, that a generic Hamiltonian system
possesses no integral (of motion) that is independent from the
Hamiltonian. However, we have the following theorem (for
finite-dimensional systems) 
\begin{thm}
Let a RIS $(\tilde{\Gamma},\tilde{\Omega},\Gamma)$ be locally reducible.
Then, $(\tilde{\Gamma},\tilde{\Omega},\Gamma)$ admits a complete system of
perennials. 
\end{thm}
A proof using Whitney embedding theorem was given in \cite{PH5}.

Two questions still arise.
\ben
\item It might be desirable that physically interesting perennials satisfy
some further conditions. For example, in field theories, one is tempted to
consider only \textit{local} functions. Are there perennials that are local?
The answer to this question, at least for general relativity, is negative
\cite{AT}.
\item Perennials, even if they exist, are likely to be complicated functions
on $\Gamma$, difficult to construct explicitly and to work with.
This may be more than just a technical problem, if we are going to quantize
the system.
\een
Let us address these two problems by studying a simple example (a
non-relativistic conservative system). Let the physical phase space be
$\mbox{\textbf{R}}^{2n}$ with canonical coordinates $q^k$ and $p_k$, $k =
1,...,n$, and let the Hamiltonian be some function $H(q,p)$ on the phase
space. We assume that the model describes some physically interesting system
so that both $q^k$ and $p_k$ are directly measurable (coordinate and
momentum). The dynamical equations are
\be
  \dot{q}^k = \{q^k,H\},\quad \dot{p}_k = \{p_k,H\}. 
\label{eq-mot}
\ee

We construct an equivalent RIS by the so-called
\textit{parametrization}. This can be done in many ways; we choose
$\tilde{\Gamma} = \mbox{\textbf{R}}^{2n +2}$ with canonical coordinates $T$,
$P$, $q^k$ and $p_k$, $k = 1,...,n$, so that $\tilde{\Omega} = dP
\wedge dT + dp_k \wedge dq^k$. Then $\Gamma$ is defined by the equation $P +
H(q,p) = 0$ and $\tilde{L_x} = L_x$ is one-dimensional,
spanned by the vector 
\[
  \left(1,0,\frac{\partial H}{\partial p_k},-\frac{\partial H}{\partial
  q^k}\right). 
\]
This form of $L_x$ implies that 1) the orbits parametrized by $T$ yield
solutions to Eqs. (\ref{eq-mot}), and 2) each $\Gamma_t$ defined by $T
= t$ is a global transversal surface.

Using global transversal surfaces, one can define a complete set of perennials
by their ``initial data'': let the perennials $Q^k_t$ and $P_{tk}$ satisfy
$Q_t^k|_{\Gamma_t} = q^k|_{\Gamma_t}$ and $P_{tk}|_{\Gamma_t} =
p_k|_{\Gamma_t}$. Then the functions $Q^k_t(T,P,q,p)$ and
$P_{tk}(T,P,q,p)$ are well-defined on $\Gamma$, because they must be
constant along the c-orbits. This particular kind of perennials were
well-known to Bergmann, who called them ``canonical constants of motion of the
Hamilton-Jacobi theory'' \cite{bergmann1}. As functions of $t$, they are 
special cases of Rovelli's ``evolving constants of motion'' \cite{rovelli}.

Let us list some properties of these perennials. First, Kucha\v{r} has
given an important argument \cite{KK2} that these perennials are not
directly observable (that is, they are in principle not measurable in
the quantum mechanics) at a time different from $T = t$. Second, at $T
= t$, the values of these perennials coincide with values of
observable quantities, and so they are measurable. Third, the values
of these perennials are given by complicated (non-polynomial, in a
non-linear field theory surely non-local) functions on the phase space
outside of $\Gamma_t$, but their restrictions to $\Gamma_t$ are simple
(polynomial, local) functions. Observe that these perennials carry a
label that specifies the transversal surface, where they \textit{have}
a direct physical meaning and \textit{are} given by simple
functions. Thus, for a perennial to be physically reasonable, it seems
to be sufficient to require that 1) it carries a label specifying some
position, 2) the values of the perennial are measurable at the
position and 3) the functional form of the perennial at the position
is simple. This may answer the first question \cite{PH5}.

The second question can be dealth with as follows (for detail see
\cite{PH1} and \cite{PH5}). The basic knowledge necessary for any
calculation with perennials is the Poisson algebra of the perennials
(all three operations of linear combination, functional multiplication
and Poisson brackets). This algebra (e.g.\ the Poisson brackets) can
be calculated, if the functional form of all perennials in at least an
open neighbourhood in $\tilde{\Gamma}$ is known. By initial data, the
perennials are not even defined in such a neighbourhood (only at
$\Gamma$), and their form is not explicitly known outside of
$\Gamma_t$. However, the Poisson algebra of perennials has been shown
in \cite{PH1} to coincide with the Poisson algebra of their
restrictions to a transversal surface. Thus, the values of the
perennials elsewhere than at $\Gamma_t$ are not needed even for
calculations.

\section{Hamiltonian dynamics}
\label{sec:H-S}
The standard kind of time evolution of dynamical systems is based on
some additional assumptions about the structure of the corresponding
RIS.  We are going to reveal these assumptions and describe them in
symplectic-geometrical terms using the example of non-relativistic
conservative system (see Sec.\
\ref{sec:perenn}). For this system, we can of course define the surfaces
$\Gamma_t$ of constant time unambiguously: $\Gamma_t := \{x \in
\Gamma\,|\,T(x) = t\}$; they are global transversal surfaces. On each
$\Gamma_t$, we can choose the coordinates $x^k_t$ and $y_{tk}$ defined by
$x^k_t := q^k|_{\Gamma_t}$ and $y_{tk} := p_k|_{\Gamma_t}$. The pull-back
$\omega_t$ of $\tilde{\Omega}$ to $\Gamma_t$ is, then, $\omega_t = dy_{kt}
\wedge dx^k_t$ and $(\Gamma_t,\omega_t)$ is a symplectic manifold. Observe
that $(\Gamma_t,\omega_t)$ can be identified with $(\bar{\Gamma},
\bar{\Omega})$. We call these surfaces \textit{time levels}. This is
the interpretation of transversal surfaces in general \cite{PH5}.
Indeed, for example the following property of general relativity can
easily be shown. Let $\Gamma_1$ be an arbitrary transversal surface
and let $\bar{\Gamma}_1$ be the set of c-orbits that intersect
$\Gamma_1$. Let $\gamma \in \bar{\Gamma}_1$ be associated with the
spacetime $(M, g)$ that does not admit any isometry (an open dense
subset of $\bar{\Gamma}_1$). Then the intersection point $\gamma \cap
\Gamma_1$ determines a unique Cauchy surface in $(M,g)$. Thus, roughly
speaking, each transversal surface defines a particular instant of
time in each solution to Einstein's equations.

The coordinates $x^k_t$ and $y_{tk}$ represent results of measurements:
$x^k_t$ that of coordinate and $y_{tk}$ that of momentum. Given the value of
all these coordinates, a state of the system is determined; two states at
different times $T = t$ and $T = s$ are considered to be the same, if the
corresponding coordinates coincide: $x^k_t = x^k_s$ and $y_{kt} =
y_{ks}$. This is because the coordinate and momentum measurements are defined
for all times; each of them specifies a class of the ``same measurements at
different times''. This leads to a further structure in the phase space: the
system of maps $\theta_{ts} : \Gamma_s \mapsto \Gamma_t$ such that the values
of these coordinates are preserved:
\be
  x^k_t(\theta_{ts}(x_s,y_s)) = x^k_s,\quad y_{kt}(\theta_{ts}(x_s,y_s)) =
  y_{ks}.
\label{theta}
\ee
We call the maps $\theta_{ts}$ \textit{time shifts}.  Time shifts define an
equivalence relation of states: if $u \in \Gamma_t$, $v \in \Gamma_s$ and $v =
\theta_{st}(u)$, then $u$ and $v$ are equivalent---they are two states with the
same measurable properties. Thus, the maps $\theta_{st}$ satisfy the axioms:
\ben
\item $\theta_{ts}$ is a symplectic diffeomorphism of
$\Gamma_s$ to $\Gamma_t$ for all $t$ and $s$,
\item $\theta_{ts} \circ \theta_{sr} = \theta_{tr}$ for all $t$ and $s$ and
$r$,
\item $\theta_{tt} = \mbox{id}$ for all $t$.
\een
We call the set $(\Gamma_t, \theta_{ts})$ an \textit{auxiliary rest
frame} (ARF) (in general, the time index $t$ runs through a more general
index set $I$). 

On one hand, the above construction can be generalized to show that a RIS that
is obtained by parametrizing a Hamiltonian dynamical system possesses
a unique ARF's. On
the other, if a RIS possesses an ARF, then we can reduce it to a Hamiltonian
dynamical system.  Indeed, the motion of the system can then be defined as the
motion with respect of the ARF. Let us consider a particular c-orbit
$\gamma$. Define $\eta_\gamma(t)$ to be the point of intersection of $\gamma$
with $\Gamma_t$; for each $t$, and for each $\gamma$, there is an
$\eta_\gamma(t)$ and it is unique. Project $\eta_\gamma(t)$ to a fixed but
arbitrary time level $\Gamma_0$ by $\xi_\gamma(t) :=
\theta_{0t}(\eta_\gamma(t))$. We call the curve $\xi_\gamma :
\mbox{\textbf{R}} \mapsto \Gamma_0$ \textit{trajectory of the system in the
physical phase space} $(\Gamma_0,\omega_0)$. The functions (we drop the index
0) 
\[
  x^k(t) := x^k(\xi_\gamma(t)),\quad y_k(t) := y_k(\xi_\gamma(t)), 
\]
satisfy the equations \cite{PH1}
\bea
  \dot{x}^k & = & \{x^k,H(x,y)\},
\label{xdot} \\
  \dot{y}_k & = & \{y_k,H(x,y)\},
\label{ydot}
\eea
where $H(x,y)$ is independent of $\gamma$ and can be constructed directly from
the set of time shifts. In our case, it coincides with the original
Hamiltonian of the system; $\{\cdot,\cdot\}$ is the Poisson bracket of the
symplectic space $(\Gamma_0,\omega_0)$. The system
(\ref{xdot})--(\ref{ydot}) is called \textit{Schr\"{o}dinger dynamical
equation}.  The coordinates $x^k$ and $y_k$ are called
\textit{Schr\"{o}dinger observables associated with the ARF}
$(\Gamma_t,\theta_{ts})$. This is a very important point: Schr\"{o}dinger
observables are always associated with some ARF \cite{PH5}.

One can also describe Heisenberg evolution. Let $o_t$ be a perennial
that represents a measurement at the time (level) $\Gamma_t$. Then the
perennial $o_s$ defined by $o_s|_{\Gamma_s} =
o_t|_{\Gamma_t}\circ\theta_{ts}$ represents the same measurement at
the time $\Gamma_s$. The set $\{o_t|t \in
\mbox{\textbf{R}}\}$ is called \textit{Heisenberg observable associated with
the ARF} $(\Gamma_t,\theta_{ts})$. We stress again: there are no more general
Heisenberg observables; to construct or define a Heisenberg observable, an ARF
is necessary \cite{PH5}. Let us project a Heisenberg observable $\{o_t|t \in
\mbox{\textbf{R}}\}$ to a particular time level $(\Gamma_0,\omega_0)$ by
$\bar{o}_t := o_t|_{\Gamma_t}$. The projection $\bar{o}_t(x,y)$ is a function
of 2$n$ + 1 variables $x^k$, $y_k$ and $t$. Its $t$-dependence satisfies the
equation \cite{PH1}
\be
  \frac{\partial}{\partial t}\bar{o}_t(x,y) + \{\bar{o}_t,H\} = 0.
\label{odot}
\ee
This is the \textit{Heisenberg dynamical equation}.

The construction of the Hamiltonian from an ARF $(\Gamma_t,
\theta_{ts})$ starts by the observation that the c-orbits define
another map $\rho_{st} : \Gamma_t \mapsto \Gamma_s$ by
$\{\rho_{st}(p)\} := \gamma_p \cap \Gamma_s$, where $\gamma_p$ is the
c-orbit through $p \in \Gamma_t$. $\rho_{st}$ is a symplectic
diffeomorphism for all $t$ and $s$ \cite{PH1}. Then the map $\chi_t :
\Gamma_0 \mapsto \Gamma_0$ defined by $\chi_t := \theta_{0t} \circ
\rho_{t0}$ is also a symplectic difeomorphism for each $t$, and its
derivative $d_t\chi_t$ defines a locally Hamiltonian vector field on
$\Gamma_0$. The corresponding Hamiltonian function $H^S_t$ is called
\textit{Schr\"{o}dinger Hamiltonian associated with the ARF}
$(\Gamma_t,\theta_{ts})$. $H^S_t$ is defined up to a constant, and
only locally in general. In general, it also depends on $t$ in a
non-trivial way (time-dependent Hamiltonian). In this case, the
Heisenberg Hamiltonian $H^H_t$ (which features in Eq.\ (\ref{odot}))
is given by $H^H_t = H^S_t \circ \chi_t$ and $H^H_t \neq H^S_t$. The
detail has been given in \cite{PH4}.

Some perennials can be exceptional in that they are associated with
measurements at more than one time level. Let $o$ be associated with
some measurement at $\Gamma_t$ and suppose that there is $s \neq t$
such that $o|_{\Gamma_s} = o|_{\Gamma_t} \circ \theta_{ts}$. Then
$o$ is also associated with the same measurement at $\Gamma_s$. A
perennial that is associated with one and the same measurement at all
time levels is called \textit{integral of motion}. For example, the
Hamiltonian of a parametrized Hamiltonian system together with all
functions which have vanishing Poisson brackets with the Hamiltonian
are integrals of motion. We can understand why integrals of motion are
exceptional. Suppose that there is a complete system of integrals of
motion. Then it must hold $\theta_{st}(\gamma \cap \Gamma_t) = \gamma
\cap \Gamma_s$ for all $\gamma$, $s$ and $t$. Such an ARF is trivial
in that the time shifts do not move the c-orbits; the dynamics is frozen.

We conclude that the perennials can represent measurements and, unlike
the standard observables, they do not need any additional structure
like ARF in order to be well-defined.

\section{Examples of ARF's}
\label{sec:ARF}
\subsection{Reduction of 2+1 gravity}
This will be very brief, for detail see \cite{moncrief}. Let us denote the
three-dimensional spacetime of the 2+1 gravity model by $(M,g)$ and the
two-dimensional Cauchy manifold by $\Sigma$. Then the points of
$\tilde{\Gamma}$ are described by pairs of fields $q_{kl}(x)$ (the
two-metric) and $\pi^{kl}(x)$ (the two-dimensional ADM momentum) on 
$\Sigma$. $\Gamma$ is determined by the three-dimensional analogon of ADM
constraints.

As time levels, Moncrief has chosen the surfaces of constant mean curvature
given by $\Gamma_\tau := \{q,\pi \in \tilde{\Gamma}\,|\,
(2\sqrt{q})^{-1}q^{kl}\pi_{kl} = \tau\}$.  ``The same states'' are defined as
those which have the same Teichm\"{u}ller parameters $q^\alpha$ and the same
values of the conjugate momenta $p_\alpha$. The knowledge of these $12h - 12$
values ($h > 1$ is the genus of $\Sigma$) determines uniquely a point at
$\Gamma_\tau$ (because the constraints are satisfied). The Hamiltonian
associated with the corresponding ARF coincides with Moncrief's
Hamiltonian. The time is one-dimensional in this case: $\tau$ runs over some
interval.

\subsection{Three forms of relativistic mechanics}
Dirac \cite{dirac2} considered a system of massive particles in Minkowski
spacetime $M$. Let us restrict ourselves just to one particle of mass
$m$. $\tilde{\Gamma} = T^*M$ with coordinates $x^\mu,p_\mu$ and
$\tilde{\Omega} = dp_\mu \wedge dx^\mu$. The constraint is $p^2 = -m^2$.

Let $G$ be the Poincar\'{e} group. The action of $G$ on $\tilde{\Gamma}$ is
defined by its 10 generator functions (via Poisson brackets) $p_\mu$
(momentum), $J_k$ (angular momentum), and $K_k$ (boost momentum). $(p_\mu
,J_k, K_k)$ forms a basis of a complete Lie algebra of perennials. Next, Dirac
chose three non-timelike surfaces in $M$ with the maximal symmetry with
respect to $G$ (``three forms''); let us take just one, the spacelike plane
$x^0 = 0$. The associated transversal surface is $\Gamma_0 := \{(x^\mu,p_\mu)
\in \tilde{\Gamma}\,|\, x^0 = 0, p^2 = - m^2\}$; define $\Gamma_g :=
g\Gamma_0$ for all $g \in G$. Thus, the time is ten-dimensional. Let
$\theta_{gh} := gh^{-1}|_{\Gamma_h}$. The Hamiltonian associated with this ARF
is $H_X(x,p) = (X^\mu p_\mu + Y^kJ_k + Z^kK_k)|_{\Gamma_0}$, where $X =
(X^\mu,Y^k,Z^k)$ is an element of the Lie algebra of $G$. For more detail, see
\cite{PH1} and \cite{PH0}.

This is an elegant construction of dynamics, where the complete system of
perennials generates a group and contains all Hamiltonians. Observe that the
maps $\theta_{gh}$ can be extended to symplectic diffeomorphisms on the whole
of $\tilde{\Gamma}$. The method is applicable to less trivial systems than the
relativistic particle. An example seems to be the torus sector of the 2+1
gravity, for which one might choose the group ISO(2,1) to play the role of the
Poincar\'{e} group (\cite{ziewer}). Its generators that form a complete Lie
algebra of perennials have been found by Moncrief \cite{moncrief2}.

Being confronted with a whole family of Hamiltonians, one can wonder
if the condition of positivity applies to them, and if not, which form
will this condition take. Indeed, there is no reason for boosts, space
shifts or rotations to be positive, and they are not. However, in
constructing the quantum theory by the so-called group quantization
method (see \cite{ishamG}), one looks for some suitable representation
of the group generated by the Lie algebra of perennials (Poincar\'{e}
group) \cite{PH1}, and one can require that in this representation the
spectrum of $p_\mu$ lies in the future light cone. Thus, the
positivity of Hamiltonian becomes a condition on the space of states.

\subsection{Functional Schr\"{o}dinger equation}
\label{sec:schrod}
In several papers \cite{KK3}, \cite{KK1} and \cite{KK4}, Kucha\v{r} and his
co-workers have studied
models of gravity with (effectivelly) two-dimensional spacetime $M$ and
one-dimensional Cauchy manifold $\Sigma$. In all cases, they managed to find
canonical coordinates $X^\mu(x)$, $P_\mu(x)$, $q^k(x)$ and $\pi_k(x)$,
$x \in \Sigma$, in $\tilde{\Gamma}$ such that $X^\mu(x)$ are spacelike
embeddings, $X : \Sigma \mapsto M$. The constraints can then be rewritten in
the form $P_\mu(x) + H_\mu[X,q,\pi;x) = 0$. If we choose the time levels
$\Gamma_X$ by fixing the embedding $X$, then $q^k(x)$ and $\pi_k(x)$ are
canonical coordinates on $\Gamma_X$, and we can define $\theta_{X'X}$ by the
requirement that the values of these coordinates be preserved similarly to
Eq.\ (\ref{theta}). The associated Hamiltonian can then be shown to coincide
with Kucha\v{r} Hamiltonian $H_\mu([X,q,\pi;x)$ (\cite{PH4}).

Observe that the time is now $\infty$-dimensional, as $X$ is an
arbitrary spacelike embedding. Still, the set of all resulting
$\Gamma_X$ comprises only a very small subset of the set of all
transversal surfaces (this has to do with the multiple choice problem
\cite{KK5}).

\section{ARF-free dynamics}
There are RIS's that do not admit any ARF or that seem to have no
natural, unique, physically distinguished ARF (like general
relativity). One can be temted to surrender the tiring search for an
ARF. Then one should attempt to formulate the dynamics
without the help of ARF's. Such attempts may be classified as
``timeless interpretations of quantum gravity'' according to
\cite{isham}. Ours is based on the fact that not only observable
properties, but also the dynamics can be represented by
perennials. Indeed, a complete system of perennials determines all
c-orbits completely. Thus, it must be possible to extract, directly
from the definition of perennials, some small-step (local) dynamical
principle. For QFT on a background spacetime,
this turns out to be the \textit{field equation}. Let us explain how
it works.

Consider a scalar field $\phi$ of mass $m$ on a fixed globally
hyperbolic spacetime $(M,g)$ with a Cauchy manifold $\Sigma$; the
field equation reads
\be
  \frac{1}{\sqrt{|g|}}\partial_\mu(\sqrt{|g|}g^{\mu\nu}\partial_\nu\phi) 
  - m^2\phi = 0.
\label{fe}
\ee
Such a system can be reformulated as a RIF \cite{IK}: the points of
the extended phase space $\tilde{\Gamma}$ are quadruples of fields, $X^\mu(x)$,
$P_\mu(x)$, $\varphi(x)$ and $\pi(x)$, on $\Sigma$; $X : \Sigma
\mapsto M$ is a spacelike embedding, $P$ is the conjugate momentum,
$\varphi := \phi|_{X(\Sigma)}$ and $\pi := (\sqrt{\gamma} n^\mu
\partial_\mu \phi)|_{X(\Sigma)}$, where $\gamma_{kl}$ is the
metric induced on $X(\Sigma)$ by $g_{\mu\nu}$ and $n^\mu$ is the unite
normal vector to $X(\Sigma)$ in $M$. The constaint surface $\Gamma$ is
determined by an equation of the form (see \cite{IK}) 
\be
  P_\mu(x) + H_\mu[X,\varphi,\pi;x) = 0.
\label{c}
\ee
Thus, points of $\Gamma$ can be labelled by the triples
$(X,\varphi,\pi)$. 

A fixed embedding $Y$ defines a transversal surface $\Gamma_Y$ by
$\Gamma_Y := \{(X,\varphi,\pi) \in \Gamma\,|\, X = Y\}$; $\varphi,\pi$
are coordinates on the physical space $\Gamma_Y$ and there is an ARF
$(\Gamma_X,\theta_{XY})$ analogous to that in Sec.\ \ref{sec:schrod}.
One can try to quantize the system by the functional Schr\"{o}dinger
equation method; the main obstacle are anomalies. A thorough
study of a QFT model on a 2-dimensional spacetime in \cite{KK6} shows how
the anomalies can be neutralized; the method has not yet been extended to
higher dimensional spacetimes, however. Anomalies may set limits to
enlargements of ARF's in general.

QFT on a general background (in the non-parametrized version) has been
intensively studied in the seventies and eighties, and it seems to be
well-understood today (for a review, see \cite{wald}). In particular,
if the background metric does not admit any timelike Killing vector,
then the quantum field algebra has no unique physical (Hilbert space)
representation and there is no unique Hamiltonian. Thus, it may be
futile to look for some Hilbert space of states and for a
Schr\"{o}dinger-like evolution in this Hilbert space.  Two methods of
how this problem can be met have been worked out: the scattering and
the algebraic methods \cite{wald}. These methods find a very natural
reformulation in terms of perennials \cite{HI2}.

\subsection{Scattering method}
\label{sec:scatt}
Suppose that there are two Cauchy surfaces, $X_1(\Sigma)$ and
$X_2(\Sigma)$ in $M$ such that $X_2(\Sigma)$ lies in the future of
$X_1(\Sigma)$ and each $X_i(\Sigma)$ possesses a static neighbourhood
$U_i$ in $M$, $i = 1,2$ (this can be generalized to include also the
scattering between two asymptotic regions). Then, one can choose a
positive-frequence basis $\{\Psi_{i\omega}\}$ for the field in $U_i$,
the frequence $\omega$ being defined with respect to the Killing
vector. The functions $\Psi_{i\omega}$ are interpreted as wave
functions of particles of $(i,\omega)$-kind; the Fock spaces $F_i$
constructed from them are called \textit{in} and \textit{out Hilbert
spaces}. As the $\omega$-spectra of in and out particles may be
different, the two Fock spaces cannot in general be identified, and we
have no unique physical Hilbert space.

One possible choice of perennials is the following
\cite{HI2}. Any point $(\varphi,\pi)$ of $\Gamma_{X_i}$ can be
considered as an initial datum for the field equation (\ref{fe}) and
the solution can be expanded in the basis $\{\Psi_{i\omega}\}$; the
coefficients $a_{i\omega}$ and $a^*_{i\omega}$ of this expansion are,
therefore, functions $a_{i\omega}(\varphi,\pi)$ on the physical phase
space $\Gamma_{X_i}$. $a_{1\omega}(\varphi,\pi)$ can then be
considered as the initial datum for a perennial, $A_{1\omega}$, on
$\Gamma$. The value of $A_{1\omega}$ at $(X,\varphi,\pi) \in \Gamma$
can be calculated as follows. The pair $(\varphi,\pi)$ defines an
initial datum along $X(\Sigma)$ for the field equation; let us denote
the corresponding solution by $\phi$. $\phi$ defines, in turn, an
initial datum $(\varphi_1,\pi_1)$ at $X_1(\Sigma)$. Then,
$A_{1\omega}(X,\varphi,\pi) = a_{1\omega}(\varphi_1,\pi_1)$.
Analogously, the perennial $A_{2\omega}$ is defined. The Lie algebra of
the complete system $\{A_{1\omega}\}$ ($\{A_{2\omega}\}$) of
perennials is called \textit{in} (\textit{out}) \textit{algebra}.

As the field equation is linear, the relation between the in and out
algebras is linear: 
\be
A_{1\omega} = \sum_{\omega'}
(\alpha_{\omega\omega'} A_{2\omega'} + \beta_{\omega\omega'}
A^*_{2\omega'});
\label{bog}
\ee 
from the Bogoliubov coefficients $\alpha$ and $\beta$,
the scattering matrix can be calculated \cite{wald}.

The construction above shows the role of the field equation in the
definition of perennials. In a linear theory like QFT on background,
the classical field equation can be directly promoted to a
\textit{quantum field equation}: just replace the real field $\phi$
by a quantum field $\hat{\phi}$ (an operator valued distribution). The
relation between the quantum in and out algebras
$\{\hat{A}_{1\omega}\}$ ($\{\hat{A}_{2\omega}\}$) obtained from the
quantum field equation has then the same form as (\ref{bog}). Observe
that the quantum field equation is less awkward and more 4-covariant
than the Wheeler-DeWitt equation based on (\ref{c}) (functional
Schr\"{o}dinger equation).

Of course, interesting models will include interaction and non-abelian gauge
symmetry, and the quantum field equation may be difficult to write
down (the products of quantum fields are not well-defined).  If
it can be written down, and if it is a reasonable dynamical equation,
then the relation between the quantum in and out algebras can be
calculated from it. This relation need not have the same form as the
corresponding classical one.

The scattering approach can be extended to a pair of transversal
surfaces $\Gamma_1$ and $\Gamma_2$ that are not global. Three
different models of this kind have been studied: two finite
dimensional systems in \cite{PH2} and \cite{PH5} and the massless
scalar field on a collapse background spacetime (Hawking effect) in
\cite{HI2}. In all cases, some c-orbit may intersect $\Gamma_1$
($\Gamma_2$) but miss $\Gamma_2$ ($\Gamma_1$). Still, the relations
between the in and out algebras are well defined and have been
calculated (for the Hawking effect, this has in fact been done in
\cite{W-hawk}). These relations cannot be, however, implemented by a
unitary map between the in and out Hilbert spaces. This happens even
in the finite-dimensional models; thus, the problem is not only 
an infinite number of out particles being created from in vacuum.

\subsection{Algebraic method}
The algebraic method of the (non-parametrized) QFT on background (see
\cite{wald}) is based on a particular class of observables, for
example those called \textit{smeared fields} $\phi_f$: each $C^\infty$
compact-support function $f: M \mapsto \mathbf{R}$ defines $\phi_f$ by
$\phi_f := \int_Md^4x\,f\phi$. Polynomials of smeared fields generate
the basic space of the algebraic approach, the so-called $C^*$
algebra; the states are certain linear functionals on this algebra,
which are interpreted as expected values; they do not form a Hilbert
space. A Hilbert space can still be obtained as a representation space
of the algebra. However, if there is no timelike Killing vector, the
algebra has no unique, physically distinguished representation. The
physical interpretation of the algebra is then based on a sufficient
amount of physically interesting states (the Hadamard quasi-free
states), the associated Hilbert spaces (GNS construction) and
well-defined expected stress-energy tensor in suitable states. For an
explanation of how such an approach to a quantum field theory works,
see \cite{wald}, Section 4.5.

In the parametrized QFT on background \cite{HI2}, the smeared fields
can be expressed as functions $\kappa_f$ on the extended phase space
$\tilde{\Gamma}$. Such functions are automatically perennials, because
they depend only on classical solutions $\phi$. The perennials
$\kappa_f$'s form a Lie algebra with the operations of
linear combination and Poisson bracket; the Poisson bracket is
determined by the field equation (\ref{fe}). In \cite{HI2}, the group
quantization method is applied to this system of perennials
\cite{ishamG}, \cite{R-group}. The resulting quantum theory coincides
with the $C^*$ algebra described above. 

The constructions described briefly in this and the foregoing
subsections lead to manifestly reparametrization invariant quantum
theories. The interpretation of these theories is based on what may be
called \textit{labelled perennials}. The labels of such perennials
carry information on the time and space \textit{position} (not
necessarily a point), where the corresponding measurement is done. If
we are given a sufficient amount of labelled perennials (like all
smeared fields, for example), then we can calculate what happens at
any position. It might, therefore, better be called
``Hamiltonian-less'' rather that ``timeless'' approach. The
``sufficient amount'' is a highly overcomplete system of perennials.
For instance, all perennials that form a complete set of Heisenberg
observables is such a system. In general, however, the labels do not
need to refer to global transversal surfaces and the perennials need
not form classes whose elements are related by time shifts (cf.\
\cite{rovelli}).

\section{Outlook}
There are some hopes that the notion of ARF will help to better
understand and to compare all current attempts at quantizing RIS's,
especially the general relativity. This requires systematic and
extended work which is currently being done; for example, a question
of how much use the different path integral methods make of
ARF is interesting. The ARF free methods may deserve further investigation.

\subsection*{Acknowledgements}
Useful discussions with A.~Ashtekar, J.~Isenberg, K.~V.~Kucha\v{r},
G.~Lavrelashvili, V.~Moncrief, C.~Rovelli, R.~M.~Wald and L.~Ziewer
are gratefully acknowledged.  The work was partially supported by the
Tomalla-Stiftung Z\"{u}rich and the Swiss Nationalfonds.


\begin{thebibliography}{99}
\bibitem{isham} C.~J.~Isham in \textit{Integrable Systems, Quantum Groups, and
  Quantum Field Theories}. Kluver Academic Publishers, London, 1993.
\bibitem{KK5} K.~V.~Kucha\v{r} in \textit{Proc.\ 4th Canadian Conference on
  General Relativity and Relativistic Astrophysics}. World Scientific,
  Singapore, 1992.
\bibitem{AM} R.~A.~Abraham and J.~E.~Marsden, \textit{Foundation of
  Mechanics}. Second Edition, Benjamin, Reading, 1978.
\bibitem{PH1} P.~H\'{a}j\'{\i}\v{c}ek, J.\ Math.\ Phys.\ \textbf{36} (1995)
  4612--4638.
\bibitem{PH5} P.~H\'{a}j\'{\i}\v{c}ek, Class.\ Quant.\ Grav.\ \textbf{13}
 (1996) 1353--1375.
\bibitem{PH4} P.~H\'{a}j\'{\i}\v{c}ek, ``Functional Schr\"{o}dinger
  Equation within a Reduction Method.'' Talk at the KK-fest, SLC, Utah,
  March 1996 (unpublished).
\bibitem{HI2} C.~J.~Isham and P.~H\'{a}j\'{\i}\v{c}ek, J.\ Math.\ Phys.\
  \textbf{37} (1996) 3522--3538.
\bibitem{PH2} P.~H\'{a}j\'{\i}\v{c}ek, J.\ Math.\ Phys.\ \textbf{36} (1995)
  4639--4666.
\bibitem{bergmann1} P.~G.~Bergmann, Rev.\ Mod.\ Phys.\ \textbf{33} (1961)
  510--514.
\bibitem{dirac1} P.~A.~M.~Dirac, \textit{Lectures on Quantum Mechanics}.
  Yeshiva University Press, New York, 1964.
\bibitem{KK2} K.~V.~Kucha\v{r} in \textit{General Relativity and Gravitation
  1992}. Ed.\ R.~J.~Gleiser, C.~N.~Kozameh and O.~M.~Moreschi. 
  Institute of Physics, Bristol, 1992.
\bibitem{AT} I.~M.~Anderson and C.~G.~Torre, Commun.\ Math.\ Phys.\
  \textbf{176} (1996) 479--539.
\bibitem{bluebook} A.~Ashtekar, \textit{Lectures on Non-Perturbative Canonical
  Gravity}. World Scientific, Singapore, 1991.
\bibitem{R-group} C.~Rovelli, Nuovo Cimento B \textbf{100} (1987) 343.
\bibitem{R-observ} C.~Rovelli in \textit{Conceptual Problems of Quantum
  Gravity}. Ed.\ by A.~Ashtekar, J.~Stachel. Birkh\"{a}user, Boston, 1991.  
\bibitem{rovelli} C.~Rovelli, Phys.\ Rev.\ \textbf{D42} (1990) 2638--2646,
  \textbf{D43} (1991) 442--456, \textbf{D44} (1991)
  1339--1341.  
\bibitem{PH0} P.~H\'{a}j\'{\i}\v{c}ek in \textit{Canonical Gravity: from
  Classical to Quantum}. Ed.\ by J.~Ehlers and H.~Friedrich. Springer, Berlin
  1994. 
\bibitem{HI1} C.~J.~Isham and P.~H\'{a}j\'{\i}\v{c}ek, J.\ Math.\ Phys.\
  \textbf{37} (1996) 3505--3521.
\bibitem{KK1} Kucha\v{r}, Phys.\ Rev.\ \textbf{50} (1994) 3961--3981.
\bibitem{AA} V.~I.~Arnold and A.~Avez, \textit{Ergodic problems of classical
  mechanics}. Addison-Wesley, Redwood City, 1989.
\bibitem{FM} A.~E.~Fischer and V.~Moncrief, Gen.\ Rel.\ Grav.\ \textbf{28}
  (1996) 207--220.
\bibitem{robinson} R.~C.~Robinson, Am.\ J.\ Math.\ \textbf{92} (19970)
  562--603, 897--906.
\bibitem{moncrief} V.~Moncrief, J.\ Math.\ Phys.\ \textbf{30} (1989)
  2907--2914. 
\bibitem{dirac2} P.~A.~M.~Dirac, Rev.\ Mod.\ Phys.\ \textbf{21} (1949) 392. 
\bibitem{ziewer} L.~Ziewer, Ph.\ D. Thesis (Berne), in preparation.
\bibitem{moncrief2} V.~Moncrief, J.\ Math.\ Phys.\ \textbf{31} (1990)
  2978--2982.
\bibitem{ishamG} C.~J.~Isham, ``Topological and Global Aspects of
  Quantum Theory,'' in \textit{Relativity, Groups and Topology II}.
  Ed.\ by B.~C.~DeWitt and R.~Stora. Elsevier, New York, 1984.
\bibitem{KK3} K.~V.~Kucha\v{r}, Phys.\ Rev.\ \textbf{D4} (1971) 955--986.
\bibitem{KK4} K.~V.~Kucha\v{r}, J.~D.~Romano and M.~Varadarajan, Dirac
  Constraint Quantization of a Dilatonic Model of Gravitational Collapse.
  Preprint, SLC, 1996; gr-qc/9608011.
\bibitem{W-hawk} R.~M.~Wald, Commun.\ Math.\ Phys.\ \textbf{45} (1975) 9.
\bibitem{IK} C.~J.~Isham and K.~V.~Kucha\v{r}, Ann.\ Phys.\ \textbf{164}
  (1985) 288.
\bibitem{KK6} K.~V.~Kucha\v{r}, Phys.\ Rev.\ \textbf{D39} (1989) 1579, 2263. 
\bibitem{wald} R.~M.~Wald, \textit{Quantum Field Theory in Curved Spacetime
  and Black Hole Thermodynamics}. University of Chicago Press,
  Chicago, 1994.


\end{thebibliography}
\end{document}